\documentclass[prb,reprint,amssymb,amsmath,numerical,superscriptaddress]{revtex4-1}
\usepackage{graphicx}
\usepackage{dcolumn}
\usepackage{bm}
\usepackage{color}
\usepackage{amssymb}
\usepackage{amsmath}
\usepackage{epstopdf}
\usepackage{bbding}
\usepackage[mathlines]{lineno}
\usepackage[dvipdfm, pdfstartview=FitH, CJKbookmarks=true, bookmarksnumbered=true, bookmarksopen=true, colorlinks, pdfborder=001, linkcolor=black, anchorcolor=blue, citecolor=blue]{hyperref}
\usepackage{lastpage}
\usepackage{fancyhdr}
\begin{document}
\title{The effect of inserted NiO layer on spin-Hall magnetoresistance in Pt/NiO/YIG heterostrucures}
\author{T. Shang}
\thanks{Present address: Swiss Light Source $\&$ Laboratory for Scientific Developments and Novel Materials, Paul Scherrer Institut, CH-5232 Villigen PSI, Switzerland}
\affiliation{Key Laboratory of Magnetic Materials and Devices $\&$ Zhejiang Province Key Laboratory of Magnetic Materials and Application
Technology, Ningbo Institute of Material Technology and Engineering, Chinese Academy of Sciences, Ningbo, Zhejiang 315201, China}
\author{H. L. Yang}
\affiliation{Key Laboratory of Magnetic Materials and Devices $\&$ Zhejiang Province Key Laboratory of Magnetic Materials and Application
Technology, Ningbo Institute of Material Technology and Engineering, Chinese Academy of Sciences, Ningbo, Zhejiang 315201, China}
\author{Q. F. Zhan}
\email{zhanqf@nimte.ac.cn}
\affiliation{Key Laboratory of Magnetic Materials and Devices $\&$ Zhejiang Province Key Laboratory of Magnetic Materials and Application
Technology, Ningbo Institute of Material Technology and Engineering, Chinese Academy of Sciences, Ningbo, Zhejiang 315201, China}
\author{Z. H. Zuo}
\affiliation{Key Laboratory of Magnetic Materials and Devices $\&$ Zhejiang Province Key Laboratory of Magnetic Materials and Application
Technology, Ningbo Institute of Material Technology and Engineering, Chinese Academy of Sciences, Ningbo, Zhejiang 315201, China}
\author{Y. L. Xie}
\affiliation{Key Laboratory of Magnetic Materials and Devices $\&$ Zhejiang Province Key Laboratory of Magnetic Materials and Application
Technology, Ningbo Institute of Material Technology and Engineering, Chinese Academy of Sciences, Ningbo, Zhejiang 315201, China}
\author{L. P. Liu}
\affiliation{Key Laboratory of Magnetic Materials and Devices $\&$ Zhejiang Province Key Laboratory of Magnetic Materials and Application
Technology, Ningbo Institute of Material Technology and Engineering, Chinese Academy of Sciences, Ningbo, Zhejiang 315201, China}
\author{S. L. Zhang}
\affiliation{Key Laboratory of Magnetic Materials and Devices $\&$ Zhejiang Province Key Laboratory of Magnetic Materials and Application
Technology, Ningbo Institute of Material Technology and Engineering, Chinese Academy of Sciences, Ningbo, Zhejiang 315201, China}
\author{Y. Zhang}
\affiliation{Key Laboratory of Magnetic Materials and Devices $\&$ Zhejiang Province Key Laboratory of Magnetic Materials and Application
Technology, Ningbo Institute of Material Technology and Engineering, Chinese Academy of Sciences, Ningbo, Zhejiang 315201, China}
\author{H. H. Li}
\affiliation{Key Laboratory of Magnetic Materials and Devices $\&$ Zhejiang Province Key Laboratory of Magnetic Materials and Application
Technology, Ningbo Institute of Material Technology and Engineering, Chinese Academy of Sciences, Ningbo, Zhejiang 315201, China}
\author{B. M. Wang}
\affiliation{Key Laboratory of Magnetic Materials and Devices $\&$ Zhejiang Province Key Laboratory of Magnetic Materials and Application
Technology, Ningbo Institute of Material Technology and Engineering, Chinese Academy of Sciences, Ningbo, Zhejiang 315201, China}
\author{Y. H. Wu}
\affiliation{Department of Electrical and Computer Engineering, National University of Singapore, 4 Engineering Drive 3 117583, Singapore}
\author{S. Zhang}
\email{zhangshu@email.arizona.edu}
\affiliation{Department of Physics, University of Arizona, Tucson, Arizona 85721, USA}
\author{Run-Wei Li}
\email{runweili@nimte.ac.cn}
\affiliation{Key Laboratory of Magnetic Materials and Devices $\&$ Zhejiang Province Key Laboratory of Magnetic Materials and Application
Technology, Ningbo Institute of Material Technology and Engineering, Chinese Academy of Sciences, Ningbo, Zhejiang 315201, China}
\date{\today}

\begin{abstract}
We investigate the spin-current transport through antiferromagnetic insulator (AFMI) by means of the spin-Hall magnetoressitance (SMR) over a wide temperature range in Pt/NiO/Y$_3$Fe$_5$O$_{12}$ (Pt/NiO/YIG) heterostructures. By inserting the AFMI NiO layer, the SMR dramatically decreases by decreasing the temperature down to the antiferromagnetically ordered state of NiO, which implies that the AFM order prevents rather than promotes the spin-current transport. On the other hand, the magnetic proximity effect (MPE) on induced Pt moments by YIG, which entangles with the spin-Hall effect (SHE) in Pt, can be efficiently screened, and pure SMR can be derived by insertion of NiO. The dual roles of the NiO insertion including efficiently blocking the MPE and transporting the spin current from Pt to YIG are outstanding compared with other antiferromagnetic (AFM) metal or nonmagnetic metal (NM).
\end{abstract}
\maketitle

Spin current, the motion of spin angular moment, has attracted intense interest due to the prospects of low consumption spintronic devices~\cite{maekawa2012, Wu2013}. Several experimental techniques have been developed to generate and manipulate spin current, e.g., spin pumping~\cite{Heinrich2011, Rezende2012, Kajiwara2010}, spin Seebeck effect~\cite{Uchida2008, Uchida2010, buer2012}, and also spin-Hall effect (SHE)~\cite{Hirsch1999, Wunderlich2005, Kato2004}. Recently, the generation and propagation of spin current in antiferromagnets (AFMs) including insulators or metals have been extensively investigated by various techniques~\cite{mendes2014, zhang2014, frangou2016, sinova2015, zhou2015, Seki2015,hahn2014, wang2014, wang2015, qiu2015, lin2016}. Especially, the thermally injected or dynamically pumped spin current from ferromagnetic (FM) YIG layer can flow into the NiO or CoO  AFM insulator (AFMI) layer and reach the Pt or Ta nonmagnetic metal (NM) layer where it can be converted into charge current by means of inverse spin-Hall effect (ISHE)~\cite{hahn2014, wang2014, wang2015, qiu2015, lin2016}. By inserting thin AFMI layer, the ISHE voltage is largely enhanced and exhibits nonmonotonic temperature or AFMI thickness dependence, which reaches a maximum near the N\'{e}el temperature of AFMI or with the AFMI thickness of $\thicksim$ 1-2 nm, respectively~\cite{hahn2014, wang2014, wang2015, qiu2015, lin2016}.

Several theoretical models have been proposed for the propagation of injected spin current through AFMs in NM/AFMs/FM heterostructures~\cite{cheng2014, takei2015, khymyn2015, rezende2016}. These models describe the spin-current transport and its enhancement by assuming that the AFMs are ordered at room temperatures, while the AFM ordering temperatures of thin AFMs are well below the room temperature~\cite{cheng2014, takei2015, khymyn2015, rezende2016}. How the spin current interacts with AFM order is still beyond understood. In most of these experimental or theoretical investigations, the spin current carried by spin waves, are produced in YIG layer and flow into the AFMI layer~\cite{hahn2014, wang2014, wang2015, qiu2015, lin2016,cheng2014, takei2015, khymyn2015, rezende2016}. While there is another type of spin current, which is carried by conduction electrons via SHE in strong spin-orbit coupling (SOC) NM~\cite{Hirsch1999, Wunderlich2005, Kato2004}. However, the investigations of spin-current transport from SOC NM into AFMI are rare, which can help us to further understand the mechanism of spin-current propagation in AFMs. Herein, we investigated the spin-current transport in Pt/NiO/YIG heterostructures by injecting the spin current from top Pt layer. Our results show that when approaching the antiferromagnetically ordered state of NiO, the spin-Hall magnetoresistance (SMR) amplitude is largely suppressed, implying that the AFM order prevents rather than promotes the spin-current transport.

\begin{figure}[tbp]
     \begin{center}
     \includegraphics[width=3.0in,keepaspectratio]{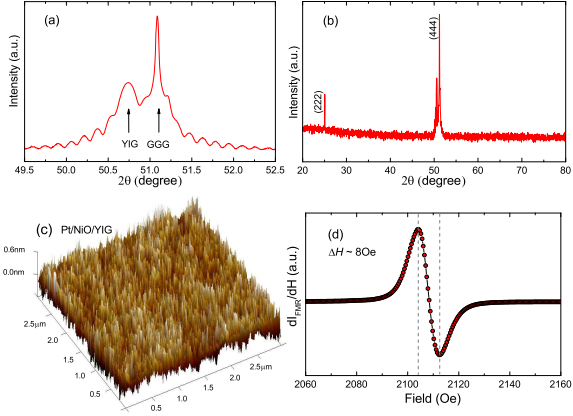}
     \end{center}
     \caption{(Color online) (a) A representative 2$\theta$-$\omega$ XRD patterns for YIG/GGG film near the (444) reflections. (b) The XRD patterns from 20 to 80 degree. (c) Three dimensional plot of the atomic force microscope surface topography for Pt/NiO(3)/YIG heterostructure over an area of 3 $\mu$m $\times$ 3 $\mu$m. (d) A representative FMR derivative absorption spectrum of YIG film with.}
     \label{fig1}
\end{figure}

The Pt/NiO/YIG heterostructures were prepared in a combined ultra-high vacuum (10$^{-9}$ Torr) pulse laser deposition (PLD) and sputter system. The high-quality YIG films were epitaxially deposited on (111)-orientated Gd$_3$Ga$_5$O$_{12}$ (GGG) substrates via PLD technique. The thin NiO films were deposited on YIG films after the growth of YIG film. The top Pt layers were sputtered in an $in$ $situ$ process. In this study, the thicknesses of YIG and Pt films are fixed at 60 nm and 3 nm, respectively, while the NiO thickness ranges from 0 to 8 nm. Figure 1(a) plot representative room-temperature x-ray diffraction (XRD) patterns for epitaxial YIG/GGG film near the (444) reflections. Clear Laue oscillations indicate the flatness and uniformity of the epitaxial films. As shown in the insets of Fig. 1(b), no indication of impurities or misorientation was detected in the range of 10 to 80 degree. The atomic force microscope surface topography of Pt/NiO(3)/YIG heterostructure over an area of 3 $\mu$m $\times$ 3 $\mu$m in Fig. 1(c) reveals a root-mean-square surface roughness of 0.14 nm (The number in the brackets represents the thickness of NiO layer in nm unit), indicating atomically flat of prepared films, with the other films showing similar surface topology.  A representative ferromagnetic resonance (FMR) derivative absorption spectrum of YIG film (60 nm) shown in Fig. 1(d) exhibits a line width $\Delta$H of 8 Oe, which was measured at radio frequency 9.39 GHz and power 0.1 mW with an in-plane magnetic field at room temperature. All the above properties indicate the excellent quality of prepared films.

\begin{figure}[tbp]
\begin{center}
\includegraphics[width=3.6in,keepaspectratio]{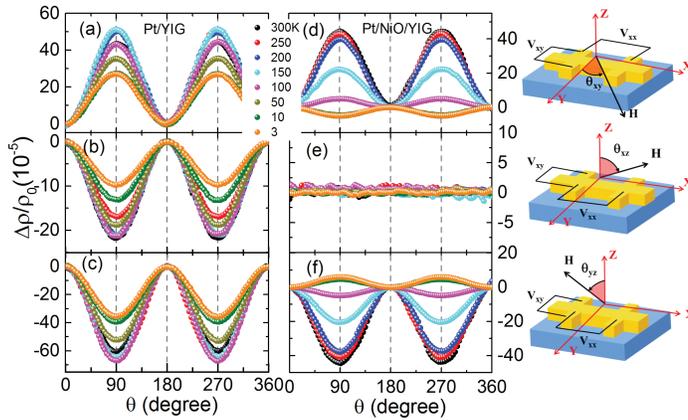}
\end{center}
\caption{Anisotropic magnetoresistance at various temperatures with the magnetic field varied within $xy$ (top-row panel), $xz$ (middle-row panel), and $yz$ (bottom-row panel) planes. The right panel shows the schematic plots of longitudinal and transverse resistance measurements. The magnetic fields are applied with angles $\theta_{xy}$, $\theta_{xz}$, and $\theta_{yz}$ relative to the $y$-, $z$-, and $z$-axes. The electric current is applied along the $x$-axis. The AMR for Pt/YIG are shown in (a)-(c), while the results of Pt/NiO(1)/YIG are shown in (d)-(f).}
\label{fig2}
\end{figure}

The recent proposed SMR built on the combined SHE and ISHE was applied to investigate the spin-current transport from SOC NM into AFMI in Pt/NiO/YIG heterostructures~\cite{sinova2015, Nakayama2013, Chen2013, Isasa2014, Chen2016}. As shown in the right panel of Fig. 2, all the films were patterned into Hall-bar geometry by using mask to measure the longitudinal and transverse Hall resistance. The anisotropic magnetoresistance (AMR) were measured in a magnetic field of 20 kOe for both Pt/NiO/YIG and Pt/NiO/MgO heterostructures. The absence of AMR in Pt/NiO/Mgo implies that the NiO moments are robust against such magnetic field. Figures 2(a)-(c) plots the AMR for Pt/YIG at various temperatures, while the results of Pt/NiO(1)/YIG are shown in Figs. 2(d)-(f). Both the Pt/YIG and Pt/NiO(1)/YIG heterostructures demonstrate clear SMR, with the amplitudes reaching 6.1 $\times$ 10$^{-4}$ and 4.5 $\times$ 10$^{-4}$ at room temperature, respectively [see Figs. 2(c) and (f)], implying that the spin current generated by SHE in Pt can transport through the NiO to interact with the YIG. Apparently, the spin current can pass through NiO from both sides, i.e., Pt $\rightarrow$ NiO $\rightarrow$ YIG or YIG $\rightarrow$ NiO $\rightarrow$ Pt~\cite{hahn2014, wang2014, lin2016}. For Pt/YIG, as shown in Fig. 2(b), the magnetic proximity effect (MPE) induced conventional AMR (CAMR) always coexists with SMR and its maximum amplitude of 2.2 $\times$ 10$^{-4}$ is comparable to the SMR. However, for Pt/NiO(1)/YIG, there is no clear CAMR even down to lowest temperature [see Fig. 2(e)], and the observed AMR are almost attributed to the SMR. As shown in Figs. 2(d)(f), the $\theta_{xy}$ and $\theta_{yz}$ scans exhibit similar behaviors, with their amplitudes being almost identical to each other. The inserted NiO layer between Pt and YIG efficiently suppresses the MPE at the interface, which also can be revealed by anomalous-Hall resistance (AHR). For instance, the AHR at 70 kOe for Pt/NiO(1)/YIG is 4.6 m$\Omega$ at 5 K, which is 6 times smaller than the Pt/YIG (27.6 m$\Omega$). As further increasing the NiO thickness, the AHR is negligible for $t_\textup{NiO} \geq$ 3 nm. Compared to the Pt/IrMn/YIG, Pt/Cu/YIG, or Pt/Au/YIG heterostructures, only 26\% of the SMR is lost by insertion of NiO, while over 80\% of the SMR is suppressed by insertion of IrMn, Cu, or Au~\cite{tian2016IrMn, Althammer2013, Chen2016}. Thus, the dual roles of the inserted NiO are outstanding, which efficiently blocks the MPE and transports the spin current.

\begin{figure}[tbp]
     \begin{center}
     \includegraphics[width=3.2in,keepaspectratio]{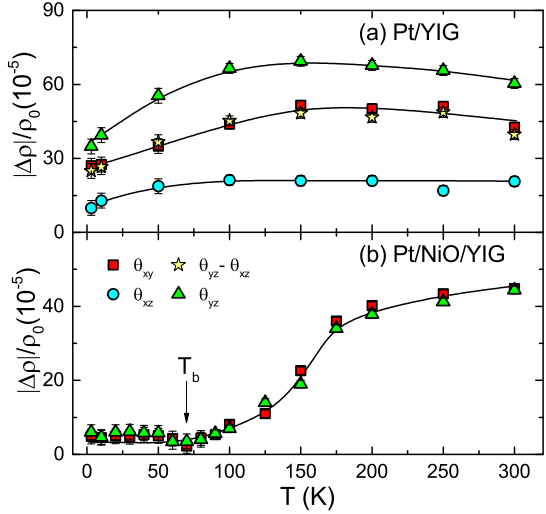}
     \end{center}
     \caption{Temperature dependence of the AMR amplitudes for (a) Pt/YIG and (b) Pt/NiO(1)/YIG heterostructures. The cubic, circle, and triangle symbols stand for $\theta_{xy}$, $\theta_{xz}$, and $\theta_{yz}$ scans, respectively. The star symbols represent the difference between SMR and CAMR ($\theta_{yz}$ - $\theta_{xz}$). The solid lines are guides to the eyes.}
     \label{fig3}
\end{figure}

All the AMR amplitudes are summarized in Fig. 3 as a function of temperature. For Pt/YIG, the SMR exhibits nonmonotonic temperature dependence and acquires its maximum value of 6.9 $\times$ 10$^{-4}$ around 150 K. While for $\theta_{xy}$ scan, where both CAMR and SMR contribute the total AMR, the amplitude is almost identical to the difference between SMR and CAMR [see star symbols in Fig. 3(a)], i.e., $\mid$$\Delta$$\rho$$\mid$/$\rho_0$($\theta_{xy}$) = $\mid$$\Delta$$\rho$$\mid$/$\rho_0$($\theta_{yz}$) - $\mid$$\Delta$$\rho$$\mid$/$\rho_0$($\theta_{xz}$), indicating that the MPE induced CAMR at Pt/YIG interface competes with the SMR. However, after inserting NiO, the SMR exhibits significantly different behaviors. As shown in Fig. 3(b), the SMR captures the AMR for $\theta_{xy}$ scan due to the absence of MPE, whose amplitude is almost identical to the $\theta_{yz}$ scan, i.e., $\mid$$\Delta$$\rho$$\mid$/$\rho_0$($\theta_{xy}$) = $\mid$$\Delta$$\rho$$\mid$/$\rho_0$($\theta_{yz}$). Similar results were also reported previously in MPE-free Rh/YIG bilayers~\cite{tianRh2015}. Moreover, the SMR undergoes a sharp decrease as decreasing the temperature and changes its sign below 70 K [see Figs. 2(d)(f)], with its amplitude being temperature independent for $T <$ 70 K. According to magnetization results, the appearance of the exchange bias field around the blocking temperature $T_\textup{b}$ $\sim$ 70 K in Pt/NiO(1)/YIG suggests that the moments in NiO layer become antiferromagnetically ordered when approaching this temperature (see details in Fig. 4). Since the sign reversal happens simultaneously around the blocking temperature, it is likely associated with the interactions between the antiferromgnetically aligned NiO moments and spin current. For Pt/NiO(1)/YIG, the SMR amplitude is almost 10 times smaller for $T <$ $T_\textup{b}$ than the room-temperature value. Most of the spin current are blocked by NiO before reaching the NiO/YIG interface when NiO moments are ordered.

\begin{figure}[tbp]
     \begin{center}
     \includegraphics[width=3.4in,keepaspectratio]{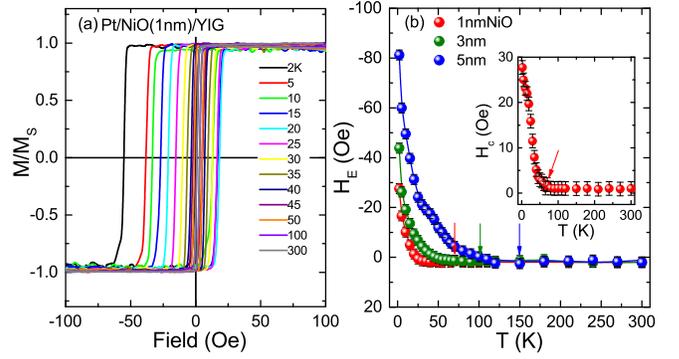}
     \end{center}
     \caption{(a) Field dependence of normalized magnetization $M$/$M_\textup{s}$ for Pt/NiO(1)/YIG at various temperatures. (b) The in-plane exchange bias field $H_\textup{E}$ for Pt/NiO/YIG with different NiO thicknesses versus temperature. The arrows indicate the AFM block temperatures $T_\textup{b}$. The inset plots the coercivity field $H_\textup{C}$ for Pt/NiO(1)/YIG.}
     \label{fig4}
\end{figure}

Since the AFM transition temperature for very thin AFMs is expected to be well below the room temperature, we also measured the field dependence of magnetization for Pt/NiO/YIG at various temperatures, from which the AFM transition temperature can be roughly tracked by blocking temperature from exchange bias field $H_\textup{E}$. As shown in Fig. 4(a), the normalized magnetic hysteresis loops $M$/$M_\textup{s}$ for Pt/NiO(1)/YIG at various temperatures are presented, with the other heterostructures showing similar behaviors. The derived $H_\textup{E}$ as a function of temperature are summarized in Fig. 4(b), from which the $T_\textup{b}$ are approximately determined to be around 70 K, 110 K, and 150 K for Pt/NiO(1)/YIG, Pt/NiO(3)/YIG, and Pt/NiO(5)/YIG, respectively. Similar blocking temperatures were previously reported in Co/NiO/(Co/Pt) heterostructures~\cite{baruth2008}. Moreover, as shown in the inset of Fig. 4(b), the coercivity $H_\textup{C}$ of Pt/Ni(1)/YIG also exhibits a step-like increase near the $T_\textup{b}$ due to the enhanced magnetic exchange coupling between NiO and YIG layer.

Different from the spin pumping or spin Seebeck techniques, where the spin current are produced at the AFM/FM interface by the precessing magnetization or thermal gradient in the YIG layer and flow through the AFMI layer~\cite{hahn2014, wang2014, wang2015, qiu2015, lin2016}, the direction of spin-current transport is reversal in our experiments. According to the SMR results, the spin current generated in Pt via SHE can transport through the NiO to interact with the YIG. However, the SMR amplitude is strongly suppressed below the blocking temperature of NiO. Based on the SMR model, the reflected spin current interact with NiO moments again before converting into charge current in Pt, which is more complicated than spin pumping or spin Seebeck process. Several theoretical models including the coherent magnetization dynamics and incoherent thermal magnons have been proposed to explain the spin-current transport in AFMI and its enhancement by insertion of AFMI ~\cite{cheng2014, takei2015, khymyn2015, rezende2016}. In the magnetic dynamics model, the processed magnetization in YIG layer exerts a torque on the AFM moments and promotes the magnetization dynamics propagation to the AFM/FM interface~\cite{cheng2014, takei2015, khymyn2015}. The other model describe the spin-current transport by considering the diffusion of incoherent thermal magnons which is caused by the accumulation of magnons at the AFM/FM interface~\cite{rezende2016}. According to these theoretical calculations, the coupling of the spin excitations at the NiO/YIG interface is much larger than at Pt/NiO or at Pt/YIG, and the spin mixing conductance of AFM/FM interface also depends on the AFMI layer thickness and linearly scales with the enhancement of spin current, both of which can explains the initial increase of spin current transport by increasing the AFMI thickness~\cite{lin2016, cheng2014, takei2015, khymyn2015, rezende2016}. In our case, the spin current is generated via SHE in Pt layer and is independent of YIG/NiO interface. However, the enhanced coupling of spin excitation at the NiO/YIG interface can efficiently enhance the spin pumping or spin Seebeck signals~\cite{cheng2014, takei2015, khymyn2015, rezende2016}. So far, all these theoretical models assume the AFMI are ordered at room temperature~\cite{cheng2014, takei2015, khymyn2015, rezende2016}, where all the simulations have been done, but the ordering temperatures of thin AFMs are well below the room temperature. Further theoretical models which include the temperature degree of freedom are highly desirable to understand the spin-current transport in AFMs.

In summary, we investigated the spin-current transport properties by means of SMR in Pt/NiO/YIG heterostructures over a wide temperature range. Different from Pt/YIG, the SMR in Pt/NiO/YIG is significantly suppressed when the temperature approaches the antiferromagnetically ordered state of NiO and becomes temperature independent below the blocking temperature. However, the dual roles of the NiO insertion including efficiently screening the MPE at Pt/YIG interface and transporting the spin current from Pt to YIG layers are outstanding. These experimental results extend the knowledge for the further theoretical investigations on the spin-current propagation in AFMs.

We thank the high magnetic field laboratory of Chinese Academy of Sciences for the FMR measurements. This work is financially supported by the National Natural Science foundation of China (Grants No. 11274321, No. 11404349, No. 51502314, No. 51522105, No. 11374312) and the Key Research Program of the Chinese Academy of Sciences (Grant No. KJZD-EW-M05). S. Zhang was partially supported by the U. S. National Science Foundation (Grant No. ECCS-1404542).

\end{document}